\documentclass[12pt]{article}

\usepackage{amssymb}
\usepackage{amsmath}

\usepackage{epsfig}
\usepackage{graphics}

\begin{document}

\title{\bf Lower bounds for open charm and beauty contributions to $\mathbf{F_2}$}

\author{A.V. Kisselev\thanks{E-mail: alexandre.kisselev@ihep.ru} \
and V.A. Petrov\thanks{E-mail: vladimir.petrov@ihep.ru} \\
\small Institute for High Energy Physics, 142281 Protvino, Russia}

\date{}

\maketitle

\thispagestyle{empty}

\bigskip

\begin{abstract}
On the basis of some peculiar scaling properties of the OPE
coefficient functions, we give the lower bounds for
$F_2^{c\bar{c}}$ and $F_2^{b\bar{b}}$ independently of the
properties of the gluon density function. Predictions for
$F_2^{b\bar{b}}$ within reach of HERA are made.
\end{abstract}

\clearpage

In deep inelastic $ep$ scattering (DIS) H1 and ZEUS Collaborations
can study an inner structure of the proton. The process with open
charm or beauty production at HERA is of particular interest,
since in these processes there is an additional (to a momentum
transfer $Q^2$) large mass scale related to a mass of a heavy
quark. Measurements of the open charm in DIS were mainly done for
$D$ and $D^*$-meson production~\cite{HERA_charm_DIS,ZEUS:04}. The
charm contribution to the proton structure, $F_2^{c\bar{c}}$, was
estimated. The $b \bar{b}$-cross section was measured in DIS by
ZEUS~\cite{ZEUS_bottom_DIS} and also in photoproduction by H1 and
ZEUS~\cite{HERA_bottom_photoproduction}. Recently, the
contribution of the beauty to the DIS structure function,
$F_2^{b\bar{b}}$, was presented by H1 in Ref.~\cite{H1:04}
(simultaneously with new data on $F_2^{c\bar{c}}$).

In our previous papers~\cite{Kisselev:97,Kisselev:03}, we have
made a description of a ratio $F_2^{c\bar{c}}/F_2$ as a function
of $x$ for different $Q^2$ values and have compared our
predictions with the HERA data. In the present paper, we calculate
lower bonds for both $F_2^{c\bar{c}}/F_2$ and $F_2^{b\bar{b}}$. We
will compare our results with the recent data from \cite{ZEUS:04}
and \cite{H1:04}, and will give predictions concerning behavior of
the ratio $F_2^{b\bar{b}}/F_2$ for not yet measured region of
$Q^2$ and $x$.

Let $\tilde{F}_2^i(Q^2,x)$ be a contribution to DIS structure
function $F_2$ from the quark of type $i$, with quark charge
squared subtracted. Then we can write
\begin{equation}\label{10}
F_2(Q^2,x) = \sum_i e_i^2 \, \tilde F_2^i(Q^2,x), \qquad i =
u,\,d,\,s,\,c,\,b.
\end{equation}
Let us define (neglecting masses of light quarks, $m_u = m_d = m_s
= 0$)
\begin{equation}\label{12}
\tilde F_2^u = \tilde F_2^d = \tilde F_2^s = \tilde F_2^q
\end{equation}
and introduce a difference between $F_2^q$ and charm (beauty)
contribution to the structure function $F_2$:
\begin{equation}\label{14}
\Delta \tilde F_2^Q = \tilde F_2^q - \tilde F_2^Q, \qquad Q =
c,\,b.
\end{equation}

At small $x$, the quantity $\tilde{F}_2^i$ can be represented in
the following form (neglecting small corrections $k^2/Q^2$ and
$m^2/Q^2$)~\cite{Kisselev:97}:
\begin{equation}\label{16}
\frac{1}{x} \, \tilde{F}_2^i (Q^2, m_i^2, x)  = \int \limits_x^1
\frac{dz}{z} \!\! \int \limits_{Q_0^2}^{Q^2(z/x)} \!
\frac{dl^2}{l^2} \,\, C \! \left( \frac{Q^2}{l^2},
\frac{m_i^2}{l^2}, \frac{x}{z} \right) \frac{\partial}{\partial
\ln l^2} \, g(l^2,z),
\end{equation}
where $ g(k^2,x)$ is a gluon density in momentum fraction $x$ at
scale $k^2$ inside the proton. As for the difference $\Delta
\tilde{F}_2^Q (Q^2, m_Q^2, x)$, we have obtained that it tends to
a $Q^2$-independent function at large $Q^2$~ (see
\cite{Kisselev:97,Kisselev:03} for details):
\begin{eqnarray} \label{18}
&& \frac{1}{x} \, \Delta \tilde{F}_2^Q (Q^2, m_Q^2, x) \Big|_{Q^2
\rightarrow \infty} \rightarrow \frac{1}{x} \, \Delta
\tilde{F}_2^Q (m_Q^2,x)
\nonumber \\
&& = \int \limits_x^1 \frac{dz}{z} \int \limits_{Q_0^2}^{\infty}
\frac{dl^2}{l^2} \,\, \Delta C \! \left( \frac{m_Q^2}{l^2},
\frac{x}{z} \right) \frac{\partial}{\partial \ln l^2} \, g(l^2,z).
\end{eqnarray}
The coefficient functions $C(u,v,x)$ and $\Delta C(v,x)$ were
analytically calculated in the first order in strong coupling
constant in Refs.~\cite{Kisselev:97}.%
\footnote{If $\Delta C$ remains analytical in $\alpha_s$ at $Q^2
\rightarrow \infty$, then it seems fairly plausible that its
scaling property holds in all orders.}
The intrinsic charm (beauty) is neglected for $Q^2 \gg Q_0^2 \gg
\Lambda_{QCD}^2$ and small $x$.

For $0 < x < 0.2$, the coefficient functions in Eqs.~\eqref{16}
and ~\eqref{18} obey the inequalities
\begin{equation}\label{20}
C(u,0,x) > \Delta C(v,x) > 0,
\end{equation}
from which one can derive the following inequalities for
{\textbf{measurable structure functions $\mathbf{F_2^{c\bar{c}}}$
and $\mathbf{F_2^{b\bar{b}}}$}~\cite{Kisselev:97}:
\begin{eqnarray}
&& \frac{F_2^{c\bar{c}} (Q^2,x)}{F_2 (Q^2,x)} > \frac{2}{5} \,
\left[1  - \frac{F_2 (m_c^2,x)}{F_2 (Q^2,x)} \right],
\label{22_1} \\
&& \frac{F_2^{b\bar{b}} (Q^2,x)}{F_2 (Q^2,x)} > \frac{1}{7}
\left[1  - \frac{F_2^{c\bar{c}} (Q^2,x)}{F_2 (Q^2,x)} - \frac{F_2
(m_b^2,x)}{F_2 (Q^2,x)} + \frac{F_2^{c\bar{c}} (m_b^2,x)}{F_2
(Q^2,x)}\right]. \label{22_2}
\end{eqnarray}
Following \cite{Martin:94}, we assumed that $F_2^{c\bar{c}}
(m_c^2,x) = F_2^{b\bar{b}} (m_b^2,x) \simeq  0$. We also neglected
a small correction $F_2^{b\bar{b}} (Q^2,x)/F_2 (Q^2,x)$ in
\eqref{22_1}, since this ratio is less than 2.5 percent even at
high values of $Q^2$~\cite{H1:04}, and, consequently, its
contribution to $F_2^{c\bar{c}} (Q^2,x)/F_2 (Q^2,x)$ is less than
$0.01$.

The formulae \eqref{16}, \eqref{18} were derived under assumption
that at small $x$ gluons make a leading contribution to a pair
production of both charm, bottom and light quarks. Let us estimate
a region of $x$ in which our formulae can be applicable. From a
NLO DGLAP fit to the proton structure function $F_2$, we know that
$g(Q^2,x)/u_{_V}(Q^2,x) \simeq 6-6.5$ for $x=0.05$, $Q^2
= 10$ GeV$^2$ (see, for instance, Fig.~3 from Ref.~\cite{PDF:exp}).%
\footnote{At smaller $x$ (and/or higher $Q^2$), a gluon dominance
is more pronounced. For instance, $g(Q^2,x)/u_{_V}(Q^2,x) \simeq
27-30$ for $x=0.01$, $Q^2 = 10$ GeV$^2$.}
Taking into account that the inequalities for the coefficient
functions \eqref{20} require $x < 0.2$, and that (schematically)
$F_2 = C \otimes g \, [x]$ ($\Delta F_2 = \Delta C \otimes g \,
[x]$ ) we get a restriction $x \lesssim 0.01$.

In our previous paper~\cite{Kisselev:03}, we estimated the lower
bound on $F_2^{c\bar{c}}/F_2$ as a function of $x$ for different
$Q^2$ values, and compared our predictions with ZEUS data on a
$D^{*\pm}$-production and the charm contribution to F2 in DIS
(taken from the fourth paper in Ref.~\cite{HERA_charm_DIS}).
Recently, the new ZEUS data on the open charm production have
appeared~\cite{ZEUS:04}. Our predictions are shown in
Figs.~\ref{fig:f2c_over_f2_1} and \ref{fig:f2c_over_f2_2} in
comparison with all ZEUS data on the $D^{*\pm}$-production.
Despite the fact that our curves give only the lower bounds for
the ratio $F_2^{c\bar{c}}/F_2$, they are very close to the
experimental points and to the ZEUS NLO QCD fit (especially, for
$Q^2 \geqslant 18$ GeV$^2$).

\begin{figure}
\resizebox{\textwidth}{!}{\includegraphics{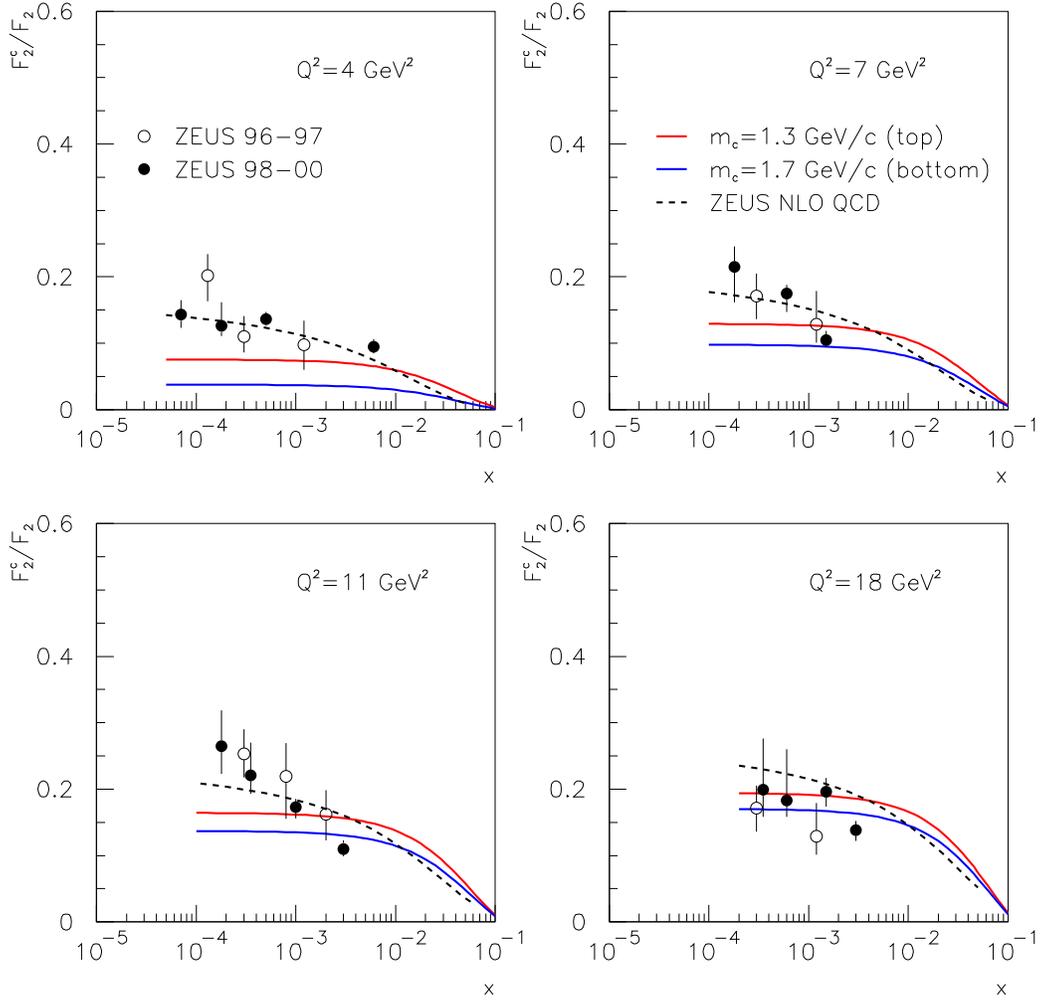}}
\caption{The ratio $F_2^{c \bar{c}}/F_2$ as a function of $x$ for
four different $Q^2$ values. Our predictions for two values of the
charm quark mass are shown by solid curves. The ZEUS data on
D*-meson production in DIS are taken
from~\cite{HERA_charm_DIS,ZEUS:04}.}
\label{fig:f2c_over_f2_1}
\end{figure}

\begin{figure}
\resizebox{\textwidth}{!}{\includegraphics{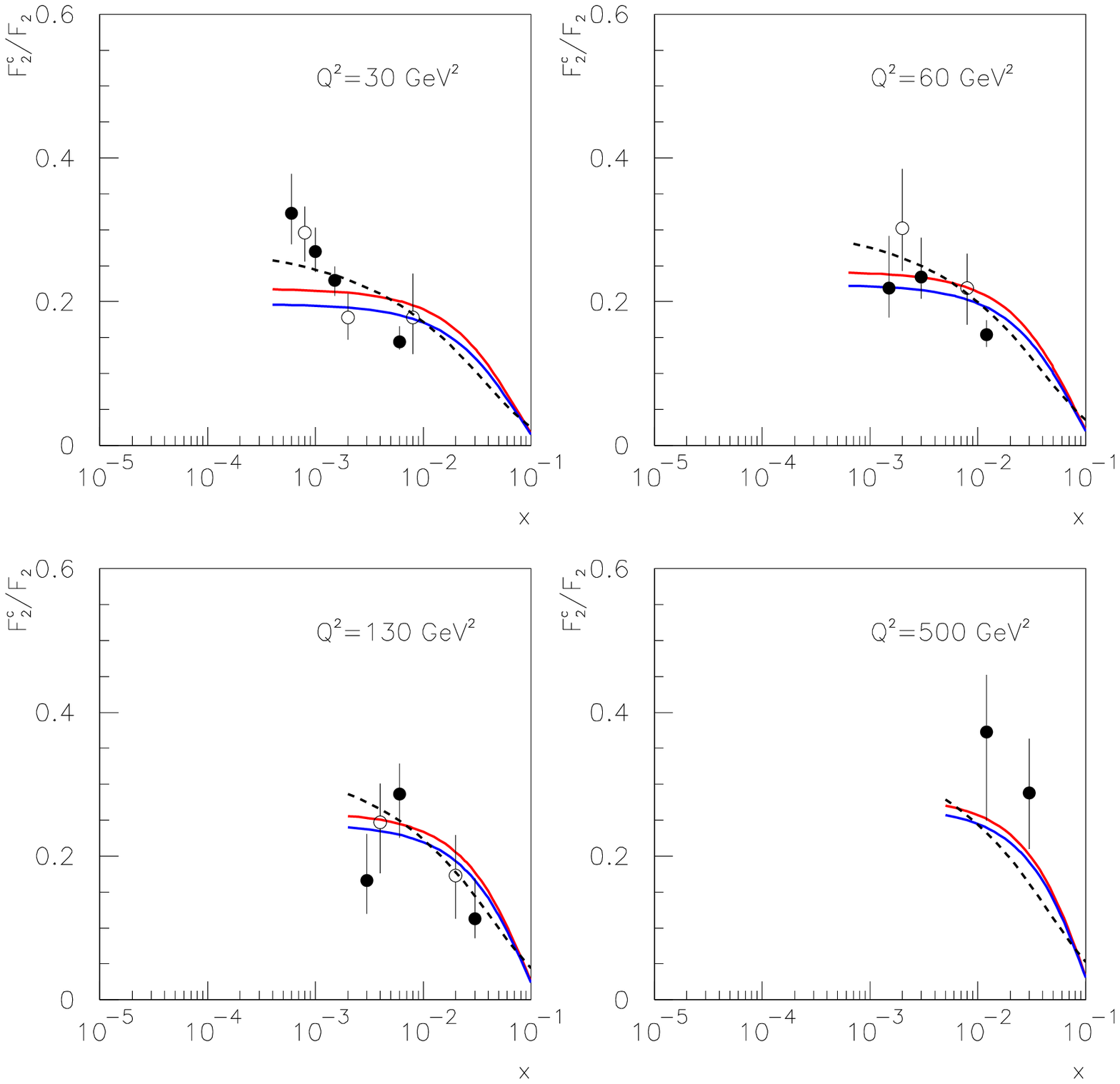}}
\caption{The same as in Fig.~\ref{fig:f2c_over_f2_1} but for
higher values of $Q^2$ .}
\label{fig:f2c_over_f2_2}
\end{figure}

Several experimental points in a region $10^{-3} < x < 10^{-2}$
lie somewhat below our curves, although coincide with the lower
bounds within $2 \sigma$. The effect may be attributed to a
contribution from valence quarks which is not negligible at not
very small $x$. In general, the model does not fix exactly
``boundary point'' value of $x$, where it is unambiguously valid.

Let us stress that in our approach \textbf{no specific expression
for the gluon distribution} $\mathbf{g(k^2,x)}$ were assumed.
Moreover, in order to calculate the lower bound for
$F_2^{c\bar{c}}/F_2$ \eqref{22_1}, one needs only the structure
function $F_2(Q^2,x)$ which has being measured with a high
accuracy. We used a parametrization of H1 from
Ref.~\cite{H1:F2fit}.

As we have already mentioned, the beauty production cross section
for DIS events has been measured by ZEUS. \cite{ZEUS_bottom_DIS},
but only recently H1 could extract $F_2^{b\bar{b}}$ as a function
of $x$ for two values of $Q^2$~\cite{H1:04}.

If we consider $F_2^{b\bar{b}}(Q^2,x)$ to be a known function of
variable $Q^2$ and $x$, we may calculate the lower bound for
$F_2^{b\bar{b}}$ by using inequality~\eqref{22_2}. In order to get
numerical predictions for $F_2^{b\bar{b}}$, we have fitted the
HERA data on open charm production~\cite{HERA_charm_DIS,ZEUS:04}
for $4 \mathrm{\ GeV}^2 \leqslant Q^2 \leqslant 500 \mathrm{\
GeV}^2$ (71 experimental points with statistical and systematic
errors added in quadrature).%
\footnote{The recent H1 data on $F_2^{c\bar{c}}$~\cite{H1:04} (4
points) were not included in the fit.}
We used a parametrization similar to that used by H1 for
$F_2(Q^2,x)$:
\begin{equation}\label{24}
F_2^{c \bar{c}}(Q^2,x) = \left[ ax^b + cx^d(1 + e \sqrt{x})\left(
\ln Q^2 + f \ln^2 Q^2 + \frac{h}{Q^2} \right) \right] (1 - x)^g.
\end{equation}
The fit gives values of parameters presented in
Table~\ref{tab:fit}, with a $\chi^2/d.\,o.f. = 55/63 \simeq 0.87$.

\begin{table}[h]
\begin{center}
\caption{The result of fitting HERA data on $F_2^{c \bar{c}}$ ($h$
in GeV$^2$)}
\bigskip
  \begin{tabular}{||c||c|c|c|c|c|c|c||}
  \hline
  a & b & c & d & e & f & g & h
  \\ \hline
  5.00 & 1.51 & 2.85 $\cdot 10^{-4}$ & -0.45 & - 3.48 & 4.38 & 2.01 &
  0.95
  \\
  \hline
  \end{tabular}
\label{tab:fit}
\end{center}
\end{table}

In Fig.~\ref{fig:f2b}, we present both the recent H1 data and a
lower bound for $F_2^{b\bar{b}}$ which was calculated with the use
of \eqref{22_2} and \eqref{24}. As in the previous case
(description of the ratio $F_2^{c\bar{c}}/F_2$), our predictions
are very closed to the experimental points.

\begin{figure}[ht]
\resizebox{\textwidth}{!}{\includegraphics{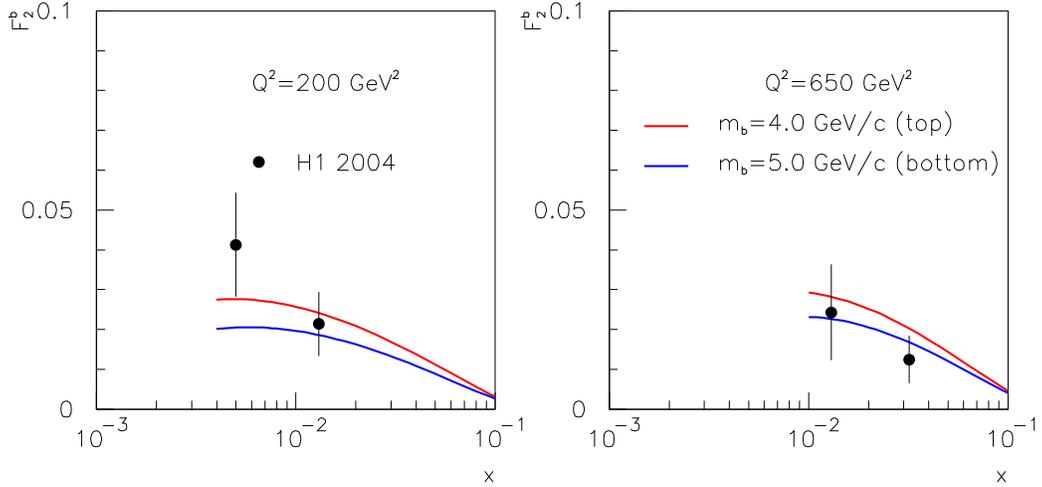}}
\caption{The beauty contribution to the structure function,
$F_2^{b \bar{b}}$, as a function of $x$ for two $Q^2$ values. Our
predictions for two values of the bottom quark mass are shown by
solid curves. The H1 data on open bottom production are taken
from~\cite{H1:04}.}
\label{fig:f2b}
\end{figure}

\begin{figure}
\resizebox{\textwidth}{!}{\includegraphics{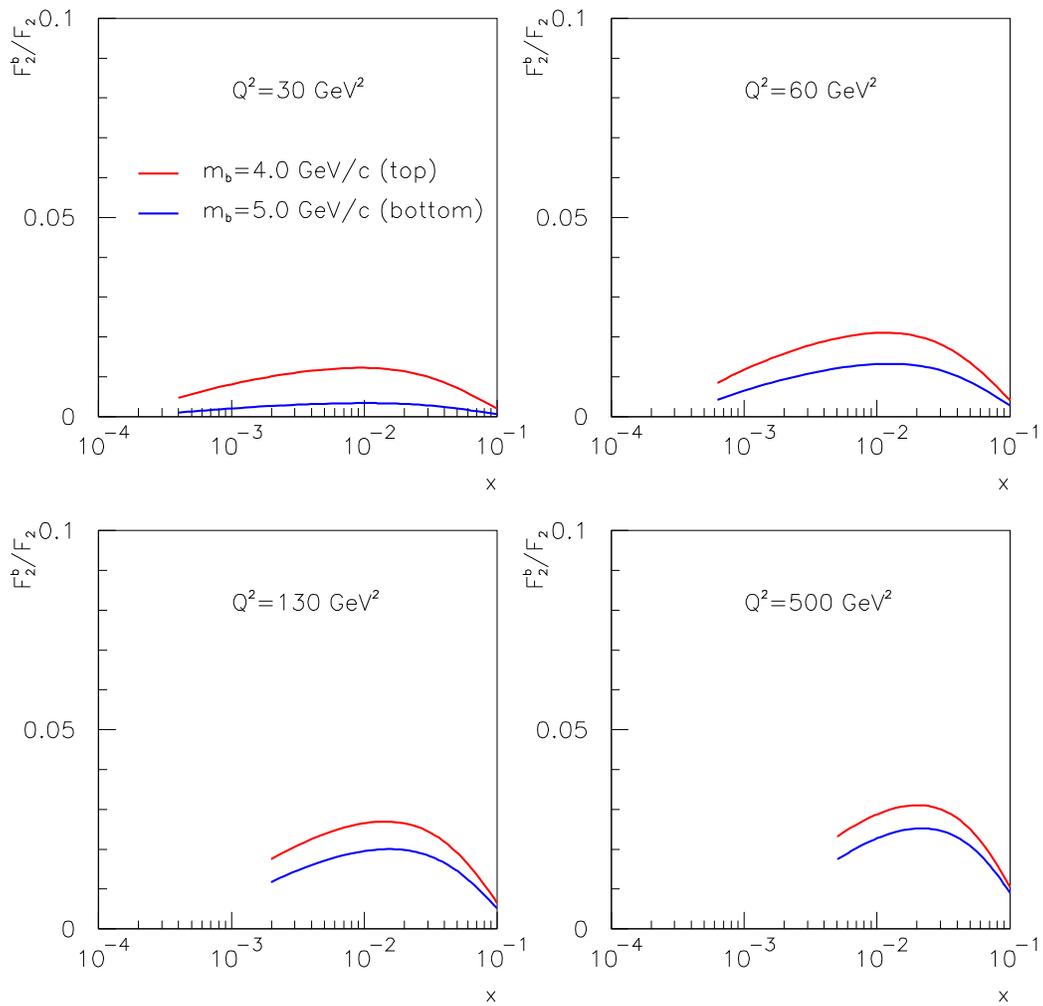}}
\caption{Our prediction for the ratio $F_2^{b \bar{b}}/F_2$. The
values of $Q^2$ and range of $x$ are taken the same as those
chosen by ZEUS in open charm measurements (see
Fig.~\ref{fig:f2c_over_f2_2}).}
\label{fig:f2b_pred}
\end{figure}

At small $x$ and large $Q^2$, theoretical predictions for the
structure functions $F_2^{c \bar{c}}$ and $F_2^{b \bar{b}}$ depend
crucially on the form of the gluon distribution $g(Q^2,x)$. There
are several fits of the parton distribution
functions~\cite{CTEQ,Alekhin:03,HERA:PDFs_fit}. To determine quark
and gluon distributions, it is necessary to make a global analysis
of a wide range of DIS and other hard processes. Nevertheless,
this procedure has a lot of experimental an theoretical
uncertainties (see \cite{Martin:04} and references therein). In
the present paper, we have given the lower bounds on $F_2^{c
\bar{c}}$ and $F_2^{b \bar{b}}$, which, however, \textbf{do not
depend on the form of $\mathbf{g(Q^2,x)}$ and on approximation
used for its calculation}. The only things we need are the
properties of the
coefficient functions%
\footnote{Let us remind that the coefficient functions $C$,
$\Delta C$ in Eqs.~\eqref{16}, \eqref{18} were calculated in order
$\mathrm{O}(\alpha_s)$, with accounting for both a gluon
virtuality and a quark mass inside a quark
loop~\cite{Kisselev:97}.}
(see Eq.~\eqref{20}) and the rise of the gluon distribution in
variable $\ln Q^2$ at $x \lesssim 0.01$, $\partial
g(Q^2,x)/\partial \ln Q^2 > 0$.

As we have already seen, the available experimental data on open
charm and beauty productions seem to saturate the lower bounds on
$F_2^{c \bar{c}}(Q^2,x)$ and $F_2^{b \bar{b}}(Q^2,x)$, at least at
$Q^2 \geqslant 18$ GeV$^2$. That is why, we expect that our
predictions for $F_2^{b \bar{b}}/F_2$ (see
Fig.~\ref{fig:f2b_pred}) will be useful for future measurements of
the structure function $F_2^{b \bar{b}}$ at HERA.


\section*{Acknowledgments}

We are grateful to L. Gladilin and M. Wing  from ZEUS
Collaboration for sending us the data on the ratio $F_2^{c
\bar{c}}/F_2$ and ZEUS NLO QCD fit.


\end{document}